\newcommand\PbNiVO{PbNi$_2$V$_2$O$_8$}
\newcommand\SrNiVO{SrNi$_2$V$_2$O$_8$}
\newcommand\PbNiMgVO%
{Pb(Ni${}_{1-x}$Mg${}_x$)${}_2$V${}_2$O$_8$}
\documentstyle[graphicx,multicol,prl,aps]{revtex}
\begin{document}
\draft
\title{Spin-Vacancy-Induced Long-Range Order in a New 
Haldane-Gap Antiferromagnet}

\author{Y. Uchiyama,$^{1}$ Y. Sasago,$^{1}$ I. Tsukada,$^{1}$ 
K.
Uchinokura,$^{1,2,}%
$\footnote{E-mail:uchinokura@ap.t.u-tokyo.ac.jp} 
A. Zheludev,$^{3}$ T. Hayashi,$^{4}$ N.
Miura,$^{4}$ and P. B\"oni$^{5}$}
\address{${}^1$Department of Applied Physics,
The University of Tokyo, 
6th Eng. Bld., Bunkyo-ku, Tokyo 113-8656, Japan}
\address{${}^2$Department of Advanced Materials 
Science, The University of Tokyo, 
6th Eng. Bld.,  Bunkyo-ku, Tokyo 113-8656, Japan}
\address{${}^3$Physics Department, Brookhaven National 
Laboratory, Upton, NY
11973-5000, U.S.A.} 
\address{${}^4$Institute for Solid State Physics, The
University of Tokyo, 7-22-1 Roppongi, Minato-ku, 
Tokyo 106-8666,
Japan} 
\address{${}^5$ Laboratory for Neutron Scattering 
ETH \& PSI, CH-5232, Villigen PSI, Switzerland}

\date{Received 12 April,  1999}
\begin{multicols}{2}[%
\maketitle
\begin{abstract}
Magnetic susceptibility, high-field magnetization and inelastic
neutron scattering experiments are used to study the magnetic
properties of a new $S=1$ quasi-1-dimensional antiferromagnet
\PbNiVO. Inter-chain interactions are shown to be almost, but not
quite, strong enough to destroy the nonmagnetic singlet ground
state and the energy gap in the magnetic excitation spectrum.
Substituting nonmagnetic Mg$^{2+}$ ($S=0$) ions for Ni$^{2+}$
($S=1$) induces a magnetically ordered state at low temperatures.
To our knowledge this is the first observation of doping-induced
long-range order in a Haldane-gap system.
\end{abstract}
]

\narrowtext


Unlike its classical counterpart, the quantum Heisenberg
antiferromagnet (HAF) has a nonmagnetic spin-singlet ground state
in one-dimensional (1D) space. 
A pioneering work by
Haldane\cite{haldane} led to a breakthrough in our understanding
of this remarkable phenomenon, and of low-dimensional magnetism in
general. Haldane suggested that the actual ground states are
totally different for integer- and half-integer-spin systems. The
half-integer 1D HAF shows quasi-long-range order (power-law spin
correlations). True long-range order can be induced by the
slightest perturbations. In contrast, for integer spins the ground
state singlet  is {\it robust}. Intrachain spin correlations decay
exponentially (true short-range order) and the excitation spectrum
has a nonzero energy gap. The system remains nonmagnetic even in
the presence of weak magnetic anisotropy or residual 3-dimensional (3D)
interactions.  The spectacular properties of the integer-spin
quantum HAF have been extensively studied theoretically and
experimentally, and are by now rather well understood. The current
issue in quantum magnetism is how the singlet ground
state is finally
destroyed and long-range order (LRO) is formed when perturbations
to the idealized Hamiltonian become sufficiently strong.

Among the factors that favor an ordered ground state are
single-ion anisotropy and  3D interchain coupling in
quasi-1D HAF systems. Increasing either of these parameters
beyond some critical level leads to a quantum phase transition to
a state with a N\'eel-like order parameter. This effect has been
extensively studied and the corresponding phase diagram is known
from numerical simulations\cite{sakai2}. 
As anticipated by Shender
and Kivelson\cite{shender}, 
another potential source of magnetic
LRO is the introduction of spin-vacancies 
in the quantum chains.
Indeed, replacing some of the  $S=1$ ions by 
$S=0$ sites severs
the chains and liberates $S=1/2$ end-chain degrees of
freedom\cite{affleck}. Under certain circumstances 
these free
spins can order in 3D space. 
To date however this
phenomenon has not been observed in any material, 
and is poorly
understood theoretically. 
In the present paper we report the first
observation of spin-vacancy-induced ordering in a new 
Haldane-gap
compound \PbNiVO, that sports a peculiar helical arrangement of
magnetic ions and substantial (yet subcritical) 
interchain interactions.

\begin{figure}
\centerline{\includegraphics*[width=8cm]{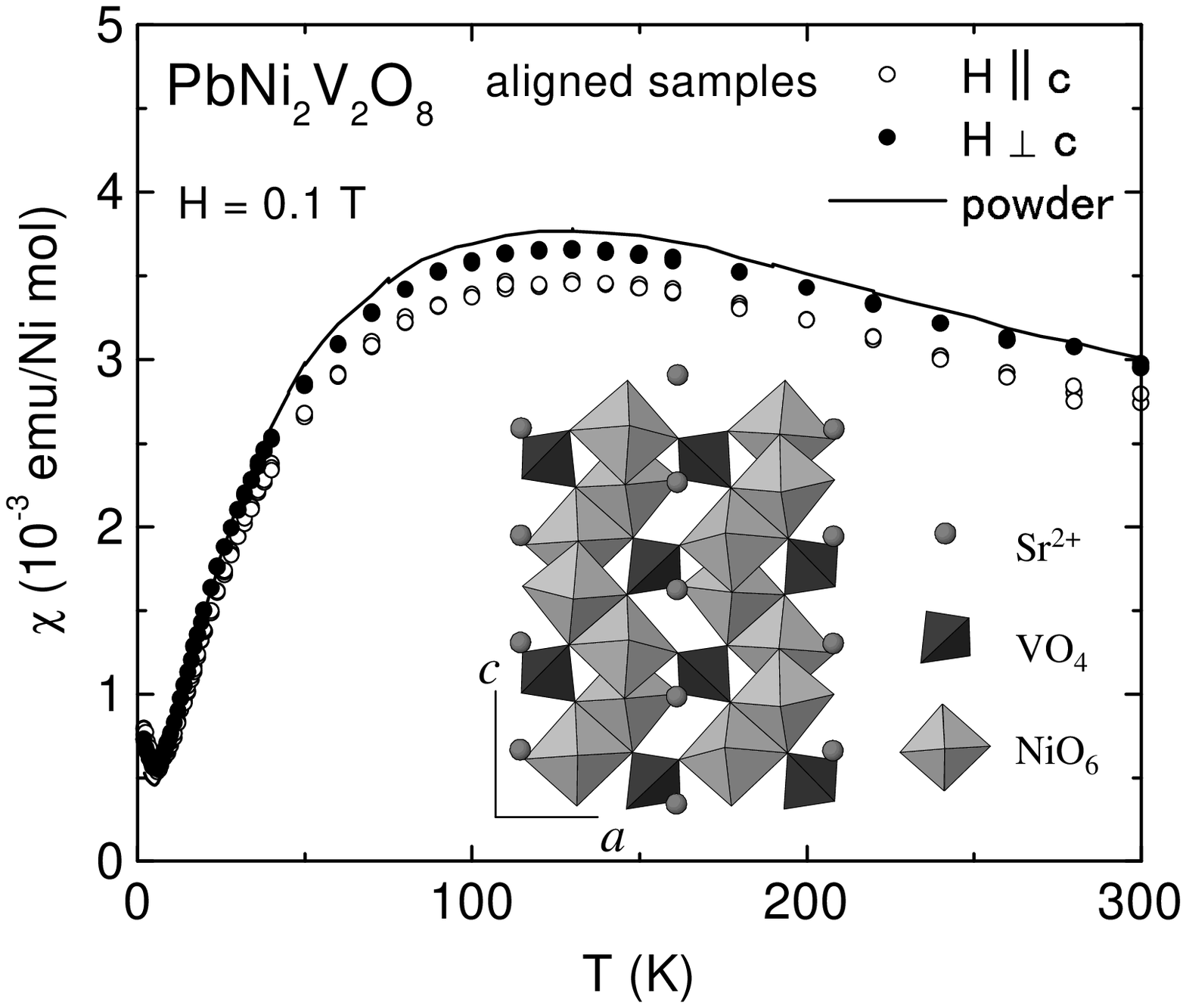}}
\caption{Temperature dependence of magnetic susceptibility
measured in PbNi$_2$V$_2$O$_8$. 
The solid line shows data for a nonaligned 
powder sample. 
The symbols are for highly aligned
polycrystalline samples. 
The magnitudes of data for aligned samples are
not exact because we cannot exactly estimate the 
quantity of the powder contained in the resin.
True magnitude for $\bbox{H}\parallel \bbox{c}$ 
should be larger than that for 
$\bbox{H}\perp \bbox{c}$ because of the $c$-axis
alignment.
Inset: A schematic view of the crystal
structure of (Sr/Pb)Ni$_2$V$_2$O$_8$.} \label{fig-st}
\end{figure}

The exact crystal structure of \PbNiVO\ has not been 
reported yet
\cite{wanklyn}. 
Preliminary X-ray diffraction measurements on
powder samples prepared by conventional 
solid-state reaction
suggest that this material is 
isomorphous with \SrNiVO\
\cite{wichmann} (Fig.~\ref{fig-st}, inset). 
\SrNiVO\ is
tetragonal (space group {\it I4${}_{\it 1}$cd}) with lattice
parameters $a$ = 12.1617 {\AA} and $c$ = 8.3247 {\AA}. We found
that \PbNiVO\ has the same crystal symmetry and very similar
lattice parameters: $a$ = 12.249(3)~{\AA} and $c$ =
8.354(2)~{\AA}. In the two isostructural compounds  slightly
distorted NiO${}_6$ octahedra are edge-shared 
around the
four-fold screw axis along the $c$ direction. All the $S=1$
Ni${}^{2+}$ ions, as well as all the nearest neighbor (NN)
Ni${}^{2+}$-Ni${}^{2+}$ bonds are equivalent. The screw-chains are
separated by VO${}_4$ tetrahedra and Sr${}^{2+}$ or Pb$^{2+}$
ions. Intrachain nearest-neighbor Ni-Ni antiferromagnetic
interactions are expected to dominate. The V$^{5+}$ sites are
presumed nonmagnetic.

\begin{figure}
\centerline{\includegraphics*[width=8cm]{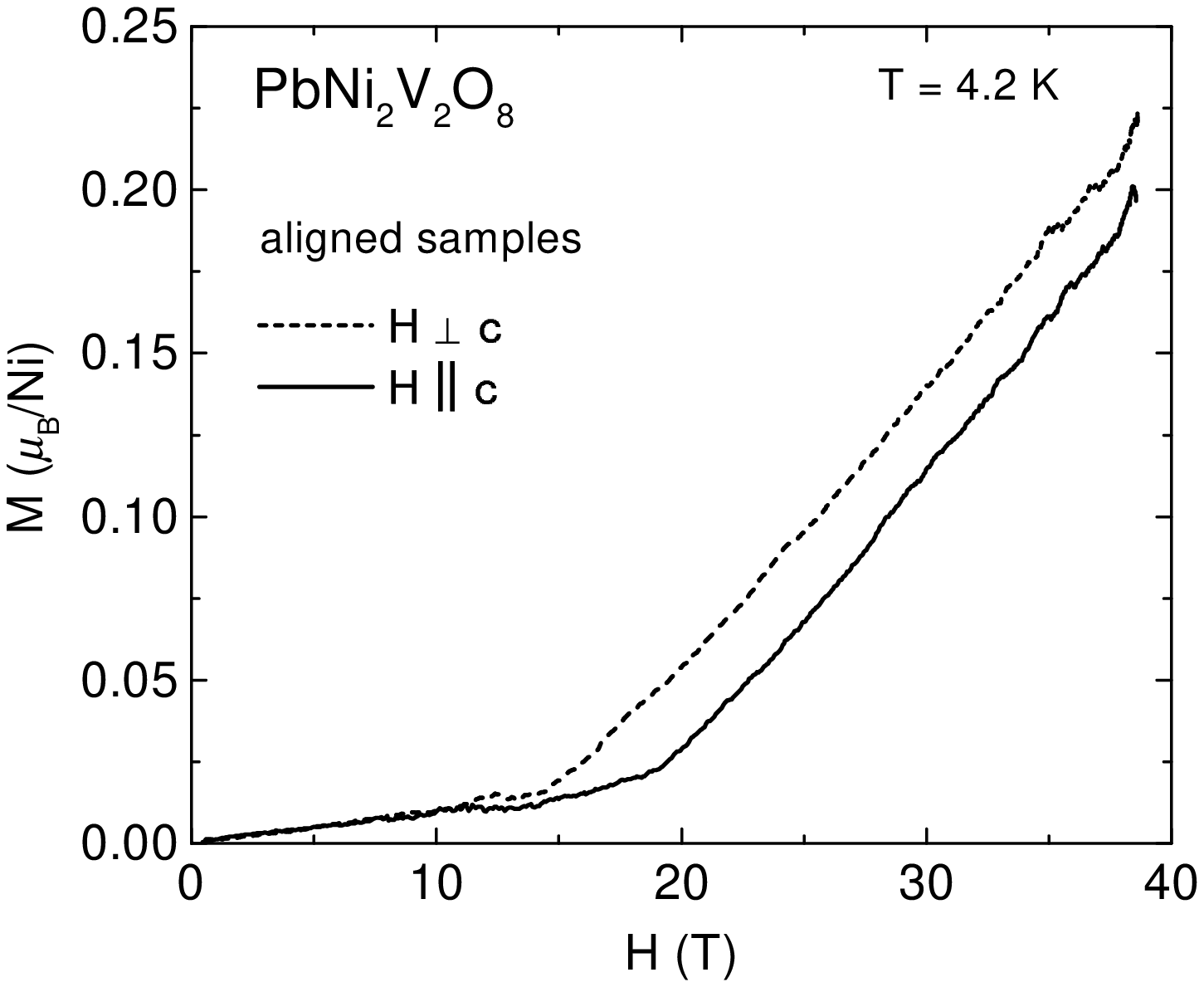}}
\caption{Magnetization curves measured in
PbNi$_2$V$_2$O$_8$ at $T=4.2$~K for magnetic fields 
parallel and
perpendicular to the chain axis.} \label{fig-mh}
\end{figure}

To characterize the new Pb-based system, bulk magnetic
measurements were performed on highly 
aligned polycrystalline
samples. 
These were prepared by applying a magnetic field of 9~T
at room temperature to a mixture of \PbNiVO\ powder and epoxy
resin (Stycast 1266). The Ni-chains tend to align themselves along
the applied magnetic field. A $c$-axis alignment of better than
3$^\circ$ (FWHM) was achieved, as determined from measuring the
rocking curve of the 004 Bragg reflection. Magnetic susceptibility
was measured with a commercial SQUID magnetometer. The
experimental $\chi(T)$ curves, shown in  Fig.~\ref{fig-st}, have a
broad maximum around 120 K for for both $\bbox{H}\perp \bbox{c}$
and $\bbox{H}\parallel \bbox{c}$. Below this temperature,
$\chi(T)$ decrease exponentially to almost zero. The residual
low-$T$ susceptibility can be attributed to the presence of
paramagnetic impurities. A straightforward analysis of the data
collected in the range $0<T<40$~K yields the thermal activation
energies: ${\Delta} = 29.4$~K $= 2.53$~meV  for $\bbox{H}
\parallel \bbox{c}$, and ${\Delta} = 27.8$~K $= 2.39$~meV for a 
transverse field, respectively.
In addition to the susceptibility measurements, 
high-field
magnetization data were taken on an aligned polycrystalline
sample in magnetic fields of up to 40~T, using the pulsed magnetic
facility at ISSP. Figure~\ref{fig-mh} shows magnetization curves
measured at $T=4.2$~K. At high fields an abrupt change of slope in
$M(H)$ is observed for both $\bbox{H}\parallel\bbox{c}$ and
$\bbox{H}\perp\bbox{c}$. The corresponding critical 
fields are
$H_c^{(\bot)}=14$~T and $H_c^{(\|)}=19$~T.

The activated temperature dependence of magnetic susceptibility is
a clear signature of a nonmagnetic ground state. The distinct
chain-arrangement of magnetic $S=1$ Ni$^{2+}$ ions in the crystal
structure suggests that we are in fact dealing with a Haldane
antiferromagnet. The behaviour of high-field magnetization is also
typical of a Haldane-gap system\cite{kazumata}. 
The starting point
for describing the magnetism of \PbNiVO\ should therefore be a
nearest-neighbor Heisenberg Hamiltonian  with a 
possible
uniaxial single-ion anisotropy (tetragonal crystal structure!): $
H = J\sum_i{\bf S}_i\cdot{\bf S}_{i+1} + D\sum_i(S_i^z)^2.$ For
not too large $D$ the ground state  is a Haldane singlet. The
lowest energy excitations are a single longitudinal 
mode and a doublet of transverse-polarized
Haldane excitations. 
Using Eq.~(2.14) in
Ref.~\cite{golinelli}, 
relating the critical fields $H_c^{(\bot)}$
and $H_c^{(\|)}$ to the doublet and singlet gap energies
$\Delta_{\bot}$ and $\Delta_{\|}$, for \PbNiVO, 
we obtain
$\Delta_{\bot}=2.2$~meV and $\Delta_{\|}=1.2$~meV. 
The intrachain
exchange constant $J$ can be deduced from the high-temperature
part of the measured $\chi(T)$ curves \cite{gadet}. For \PbNiVO\
this analysis yields $J=95$~K $=8.2$~meV. Already we see that the
mean gap energy $(2\Delta_{\bot}+\Delta_{\|})/3=1.87$~meV is
substantially smaller that the value $0.41J\approx3.36$~meV,
expected for noninteracting chains\cite{sakai2,golinelli}. 
The most
likely source of this discrepancy is substantial 
interchain
exchange interactions. In their presence the excitation energies
become dependent on the momentum transfer perpendicular to the
chain axis. $\Delta_{\bot}$ and $\Delta_{\|}$ in this case
represent the {\it global} minima in the 3D dispersion, 
and cannot
be uniquely related to $J$.

\begin{figure}
\centerline{\includegraphics*[width=7cm]{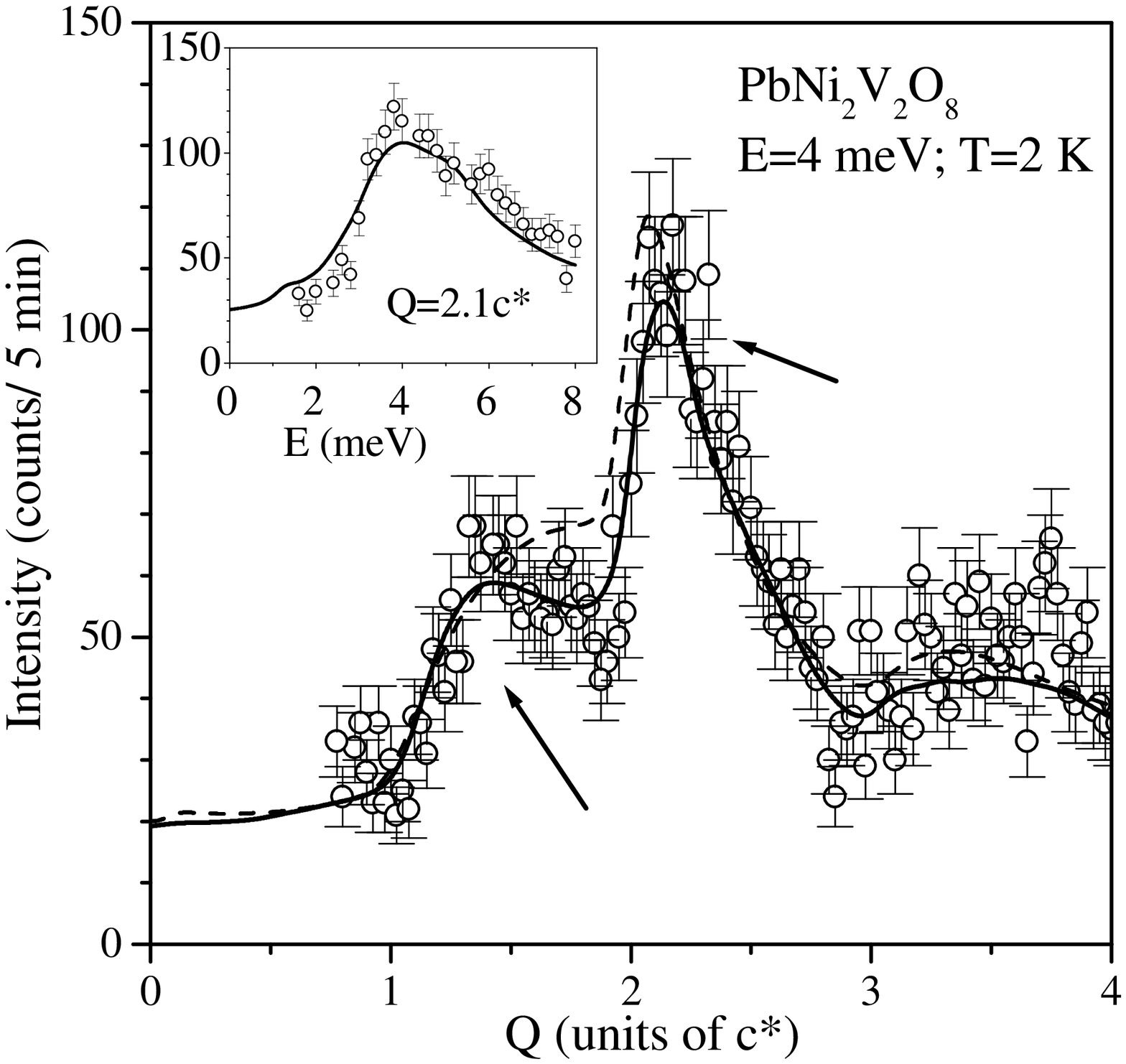}}
\caption{ Typical constant-$Q$ (inset) and 
constant-$E$ (main
panel) inelastic neutron scattering scans measured in
PbNi${}_2$V${}_2$O${}_8$ powder at $T=2$~K. Lines are model fits,
as described in the text. } \label{fig-neutron}
\end{figure}

Dynamic spin correlations in \PbNiVO\  were studied in a series of
inelastic neutron scattering experiment on a 10~g powder sample at
the Dr\"{u}chaL 3-axis spectrometer  (SINQ spallation source, PSI,
Switzerland). 
Neutrons of fixed final energy $E_f=8$~meV were
used with a (guide)-80'-80'-(open) collimation setup. An entire
series of constant-$E$ scans was collected in the energy range
2--8~meV to map out the low-energy part of the magnetic dynamic
structure factor $S(Q,\omega)$. Additional data were collected in
constant-$Q$ scans. A typical data set measured at a momentum
transfer just exceeding the 1D antiferromagnetic zone-center
$Q_{\pi}=4\pi/c$ at $T=2$~K is shown in the inset in
Fig.~\ref{fig-neutron}. This and all other const-$Q$ scans clearly
show the presence of a spin gap of roughly 3.5~meV. The main panel
in Fig.~\ref{fig-neutron} shows a constant-$E$ scan measured at
$\hbar \omega=4$~meV, i.e., just above the gap energy. In order to
analyze these data we have calculated the powder average of the
3D dynamic magnetic structure factor for an isolated
screw-shaped Haldane chain, as appropriate for \PbNiVO\
\cite{zheludev}. The relevant parameters of this cross section are
the two gap energies $\Delta_{\bot}$ and $\Delta_{\|}$. Although
this model (dashed lines in Fig.~\ref{fig-neutron}) can reproduce
the measured const-$Q$ scans fairly well and predicts the
asymmetric feature at $Q_{\pi}$, it cannot account for the sharp
peak seen at $Q\approx 3 \pi/c$ in constant-$E$ scans.

A better agreement with the experiment could only be obtained by
taking into account interchain interactions. We assumed a weak
Heisenberg coupling between nearest-neighbor Ni chains and treated
it on the RPA level. 
A detailed description of this analysis is
beyond the scope of this Letter, and will be published
elsewhere\cite{zheludev}. 
Compared to the noninteracting-chain
model only one additional parameter, 
namely the interchain
coupling constant $J_{\bot}$ is introduced. 
The total number of
variables is minimized by fixing the lowest excitation energies to
the gap values deduced from high-field measurements. 
An excellent
global fit to the data is obtained with 
$J_{\bot} =0.096\pm 0.003~\rm{meV}$
from least squares refinement
(solid lines in Fig.~\ref{fig-neutron}). 
From this value and the
measured critical fields the intrinsic gap energies 
for noninteracting chains can be determined:
$\Delta_{\bot}^{(0)}=4.1~\rm{meV}$ and
$\Delta_{\|}^{(0)}=3.7~\rm{meV}$, respectively. 
Using the
numerical results from Ref.~\cite{golinelli}, 
for the intrachain
coupling constant $J$ and the anisotropy parameter $D$ we obtain
$J=9.5$~meV and $D=-0.23$~meV (easy-axis), 
respectively. 
The
former value is in reasonably good agreement with 
the high-$T$
susceptibility estimate $J=8.2$~meV.

The results presented above enable us to place the new material on
the Sakai-Takahashi $D-J_{\bot}$ phase 
diagram\cite{sakai2}. 
For \PbNiVO\ we have $D/J\approx -0.024$, $J_{\bot}/J=0.01$ and $z=4$,
$z$ being the coordination number for interchain interactions.
Using Fig.~8 in Ref.~\cite{sakai2}, we find that 
\PbNiVO\ is
positioned in the Haldane phase, 
just below the borderline with
the ordered Ising-like phase. 
In other words, by sheer chance the
system is {\it almost critical towards the formation of long-range
order}. As will be reported elsewhere\cite{yasu}, unlike \PbNiVO,
the isostructural \SrNiVO\ system falls on the opposite side of
the transition line and actually ordered in 
3D space in a
weak-ferromagnetic structure below $T_{N}=7$~K. A simple
substitution of nonmagnetic Sr for nonmagnetic Pb slightly
alters the coupling geometry, increasing interchain 
interactions
or single-ion anisotropy beyond the critical values.

\begin{figure}
\centerline{\includegraphics*[width=8cm]{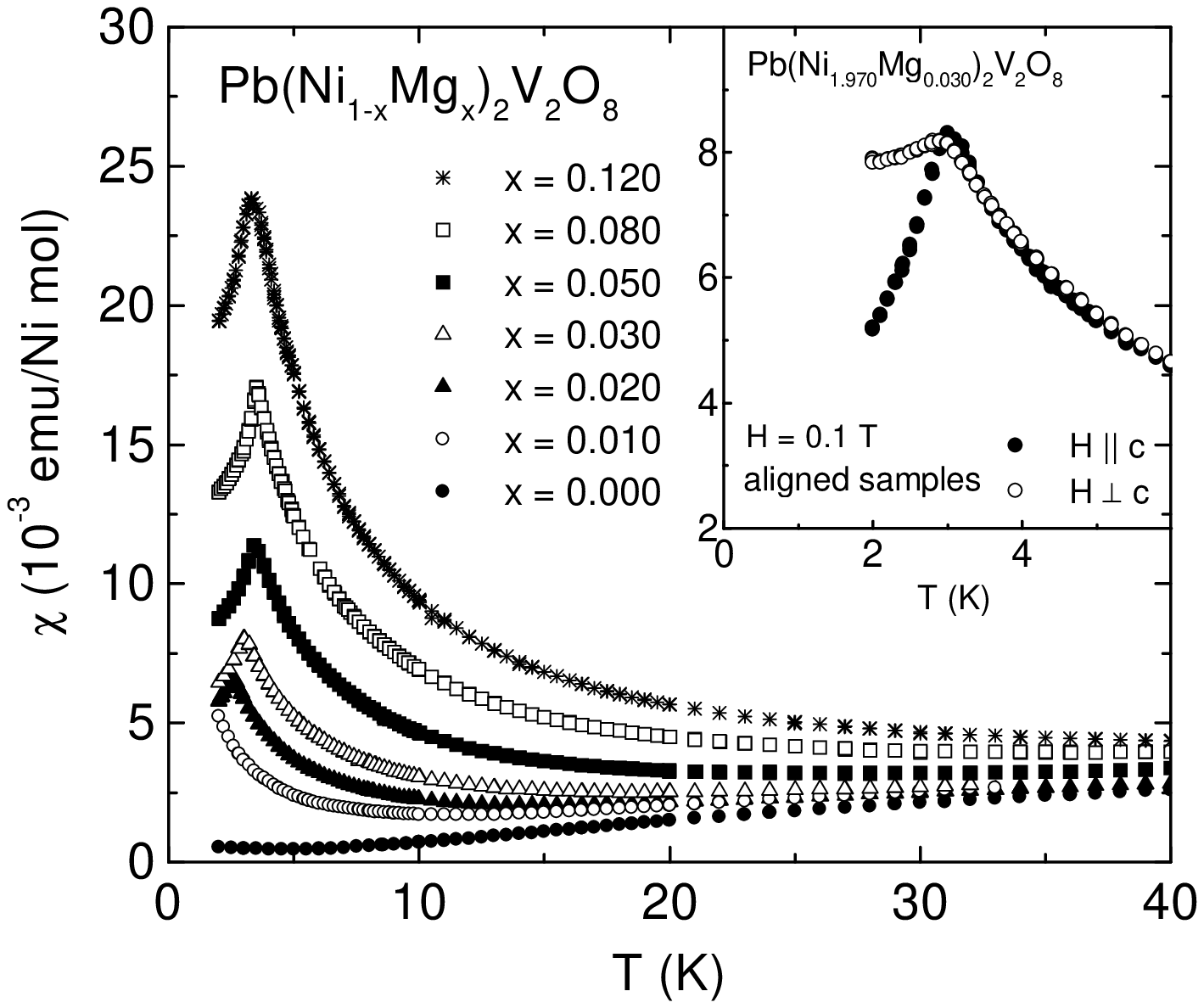}}
\caption{Temperature dependences of magnetic
susceptibility measured in Pb(Ni$_{1-x}$Mg$_x$)$_2$V$_2$O$_8$
($0\le x\le 0.120$) powder samples. The inset shows the data for a
highly aligned
Pb(Ni${}_{0.970}$Mg${}_{0.030}$)${}_2$V${}_2$O${}_8$. }
\label{fig-st-doped}
\end{figure}

Having characterized the pure \PbNiVO\ compound, we now turn to
the main subject of this paper, describing the emergence of
N\'eel-like magnetic LRO upon introduction of spin vacancies in this
Haldane-gap system. 
Nonmagnetic sites are introduced into the
spin chains by substituting $S=0$ Mg${}^{2+}$ ions for the $S=1$
Ni${}^{2+}$. The measured temperature dependences of magnetic
susceptibility for a series of Pb(Ni$_{1-x}$Mg$_x$)$_2$V$_2$O$_8$
powder samples with $x$ ranging from 0 to 0.120 are shown in
Fig.~\ref{fig-st-doped}. Except for the $x=0$ compound, an
increase of susceptibility is observed at low $T$ in all samples
studied. The enhancement scales with Mg concentration, and can be
attributed to the presence of free $S=1/2$ spins at the ends of
the Ni chains severed by the spin vacancies\cite{hagiwara}. In
this scenario the number of liberated spins is {\it twice} the
number of substitutions. To check this scaling we compared the
temperature dependence of magnetic susceptibility in 
Mg${}^{2+}$($S=0$)-
and Cu${}^{2+}$($S=1/2$)-substituted \PbNiVO. The replacement of a
Ni$^{2+}$ ion by a Cu$^{2+}$ ion also produces two end-chain
spins. However, these two spins couple antiferromagnetically 
to the Cu-spin, producing a
ground state configuration of total $S=1/2$. The total number of
added free spins in this case is {\it equal} to the total number
of added $S=1/2$ ions. Experimentally, the measured $\chi(T)$
curve for 2\%-Cu-doped \PbNiVO\ sample was found to be virtually
indistinguishable from that for a 1\%-Mg-doped sample.

The central result of this work is the observation of a sharp
decrease in the magnetic  susceptibility of
Pb(Ni$_{1-x}$Mg$_x$)$_2$V$_2$O$_8$ 
below $T_c\approx 3.5$~K for $x\ge
0.020$. The slope of $\chi(T)$ changes abruptly with reversing its
sign. This behavior is a clear indication of a magnetic phase
transition. Its nature is revealed in experiments 
on an aligned
powder sample with $x=0.030$ (Inset of Fig.~\ref{fig-st-doped}). In this
case the transition point is about $T_c=3.0$~K. Below this
temperature a large anisotropy of susceptibility is observed. For
$\bbox{H}\parallel \bbox{c}$, $\chi(T)$ decreases abruptly below
$T_c$. In contrast, for $\bbox{H}\perp \bbox{c}$ it remains
practically constant. This behavior is typical of a 
N\'eel
antiferromagnet with spins ordering along the $c$ direction, that,
as we have shown previously, is the magnetic easy axis of the
system.

The effect of spin-vacancy substitution has been studied
experimentally (see, for example, Ref. \cite{hagiwara} and
theoretically\cite{affleck}) for a number of Haldane-gap systems.
Apart from the appearance of free end-chain spins and the
resulting paramagnetic contribution to the low-$T$ susceptibility,
severing the chains leads to a reduction of the dynamic spin
correlation length. \PbNiVO\ is, to our knowledge, the first
example of a Haldane AF that orders magnetically upon doping. The
key of course is in the interchain interactions: to order
magnetically the liberated end-chain spins need to be coupled in 3D space. 
As we have demonstrated, interchain coupling is
quite substantial in \PbNiVO. 
The possibility of vacancy-induced
LRO in a Haldane-gap system was first qualitatively discussed by
Shender and Kivelson\cite{shender}. 
In particular, the authors
suggested that there is a critical impurity 
concentration below which LRO does not occur. 
As to this point more detailed low-$T$ studies are 
required.

The magnetic transition that we observe in
Pb(Ni$_{1-x}$Mg$_x$)$_2$V$_2$O$_8$ should not be confused with
that in the extensively studied Haldane system CsNiCl$_3$
\cite{buyers,kakurai}. Unlike in \PbNiVO, direct interchain
interactions in CsNiCl$_3$ are stronger than the critical value,
and even the undoped material has a magnetically ordered ground
state. 
In fact, the introduction of Mg$^{2+}$ vacancies leads to a
{\it decrease} of the ordering temperature \cite{sosin}, in
striking contrast with the behavior of \PbNiVO. Rather, the
situation in Mg-doped \PbNiVO\ is similar to that in the
spin-Peierls CuGeO$_3$, doped with Zn\cite{Hase}, Mg\cite{masuda}
or Ni\cite{koide}, and  the two-leg ladder system
SrCu${}_2$O${}_3$ doped with Zn\cite{azuma}. Although these
materials are half-integer-spin antiferromagnets, their ground
states fall in the same universality class as the Haldane state,
and they both have a gapped magnetic excitation spectrum. Just as
we propose for \PbNiVO, the onset of long-range order is believed
to be caused by interchain interactions that couple free spins
liberated by impurity substitution. Doping-induced long-range
magnetic order thus appears to be a {\it universal} feature of
quasi-1D singlet quantum antiferromagnets with 
the excitation gap.

In summary, we have shown that \PbNiVO\ is a Haldane-gap system
with substantial interchain interactions and is almost critical
towards the formation of long-range order. For the first time
long-range magnetic ordering was induced in a Haldane
antiferromagnet by spin-vacancy substitution on the $S=1$ sites.

We would like to thank T.~Masuda for the coorperation in the
various aspects of this work. This work was supported in part by
Grant-in-Aid for COE Research ``Phase Control of
Spin-Charge-Photon (SCP) Coupled System'' from the Ministry of
Education, Science, Sports, and Culture. Work at Brookhaven
National Laboratory was carried out under Contract
   No. DE-AC02-98CH10886,
   Division of Material Science, U.S. Department of Energy.

\end{multicols}
\end{document}